\documentclass[12pt]{article}
\usepackage{a4wide}
\usepackage{graphicx}
\usepackage{amssymb}

\newcommand{\be}{\begin{equation}}
\newcommand{\ee}{\end{equation}}
\newcommand{\ba}{\begin{eqnarray}}
\newcommand{\ea}{\end{eqnarray}}
\newcommand{\treeC}[1] {\left[#1\right]_{C_{i}}} 

\begin{document}

\begin{titlepage}
\begin{flushright}
LU TP 09-17\\
arXiv:0906.3118 [hep-ph]\\
revised August 2009
\end{flushright}
\vfill
\begin{center}
{\Large\bf Relations at Order $p^6$ in Chiral Perturbation Theory}
\vfill
{\bf Johan Bijnens and Ilaria Jemos}\\[0.3cm]
{Department of Theoretical Physics, Lund University,\\
S\"olvegatan 14A, SE 223-62 Lund, Sweden}
\end{center}
\vfill
\begin{abstract}
We report on a search of relations valid at order $p^6$ in
Chiral Perturbation Theory. We have found relations between $\pi\pi$,
$\pi K$ scattering, $K_{\ell4}$ decays, masses and decay constants
and scalar and vector form factors. In this paper we give the relations
and a first numerical check of them.
\end{abstract}
\vfill
\vfill
\end{titlepage}

\section{Introduction}

Chiral Perturbation Theory (ChPT) \cite{Weinberg0,GL0,GL1} is the effective
field theory for the strong interaction at low energies. Some recent reviews
are \cite{review,reviewp6,EFT09hans}. In the mesonic sector many calculations
have now been performed to two-loop or next-to-next-to-leading order (NNLO),
 see
the review \cite{reviewp6}. Since in an effective field theory like ChPT there
appear new Lagrangians at every order, tests  of ChPT at NNLO are difficult to
perform since for most processes new combinations of these parameters,
called low-energy constants (LECs), appear.

One way to test ChPT at NNLO order is to find observables where the same
combinations of LECs appear. Many of these pairs of observables were found
in the explicit calculations but no systematic study had been done. That is
the purpose of this work. We take systematically all observables that
can contain a dependence on the NNLO LECs in $\pi\pi$ and $\pi K$ scattering,
the masses and decay constants, $\eta\to3\pi$, $K_{\ell4}$ and the scalar and
vector formfactors and determine how many of these contain the same
combinations of NNLO LECs. Of the 76 observables we include, we find 35
such combinations. These are discussed in Sects.~\ref{pipi} to
\ref{finalrelation}. These allow in principle to test the validity of three
flavour ChPT at NNLO. However, many relations involve poorly known quantities
from the scalar formfactors so we have restricted the numerical discussion
to the $\pi\pi$, $\pi K$ and $K_{\ell4}$ sector. The tests in the vector
formfactors were already discussed extensively in the earlier work, so
we do not present numerical results for those either.
We find a mixed picture. Three flavour ChPT mostly works but
 there are problems.
Some preliminary results were presented in \cite{EFT09ilaria}.

We first discuss the $\pi\pi$ threshold parameters relations in both two and
three flavour ChPT in Sect.~\ref{pipi}. After that we restrict ourselves
to three flavour ChPT, first $\pi K$  threshold parameters relations in
Sect.~\ref{piK} and the relations between both sectors in Sect.~\ref{pipipiK}.
Sect.~\ref{Kl4} discusses the relation between $K_{\ell4}$ and $\pi K$
scattering. For $\eta\to3\pi$ we find relations involving the cubic
dependence of the Dalitz plot in Sect.~\ref{e3pi}. For the scalar formfactors
we find the known relations and one new one, Sect.~\ref{scalarFF}, but when
relating the scalar sector to other sectors we find several new relations as
discussed in Sects.~\ref{scalarFFmass}, \ref{vector}, \ref{scalarFFpipipiK}
and \ref{finalrelation}. We shortly recapitulate our conclusions in
Sect.~\ref{conclusions}. For completeness, we have added in an appendix
the correspondence between the subthreshold and threshold parameters
for $\pi\pi$ and $\pi K$ scattering.

\section{Notation}

The Lagrangian at NNLO contains 90 LECs, called the $C_i$ in \cite{BCE1,BCE2}.
Since in this work we check whether the same combinations of LECs appear we
use the notation
\be
\treeC{A} = C_i^r\textrm{-dependent~part~of~}A\,.
\ee
We also use the notation $B_0$ and $F_0$, the chiral limit of $F_\pi$, the two
constants that appear in the lowest order Lagrangian \cite{GL1}.
For the physical observables we use for each case the established notation.

We always express dimensionful quantities in the appropriate units of
$m_{\pi^+}$, which is in any case standard practice for many of the quantities
we consider. We use the symbol $\rho=m_K/m_\pi$ to indicate the kaon mass.
This way the relations are easier to write down.

\section{$\pi\pi$ scattering}
\label{pipi}

The $\pi\pi$ scattering amplitude can be written as a function $A(s,t,u)$
which is symmetric in the last two arguments:
\begin{equation}
A(\pi^{a}\pi^{b}\rightarrow\pi^{c}\pi^{d})=
\delta^{a,b}\delta^{c,d}A(s,t,u)+\delta^{c,d}\delta^{b,d}A(t,u,s)
+\delta^{a,d}\delta^{b,c}A(u,t,s)\,,
\end{equation}
where $s,t,u$ are the usual Mandelstam variables.
 The isospin amplitudes $T^{I}(s,t)$ $(I=0,1,2)$ are
\ba
T^{0}(s,t)&=&3A(s,t,u)+A(t,u,s)+A(u,s,t)\,,
\nonumber\\
T^{1}(s,t)&=&A(s,t,u)-A(u,s,t)\,,
\nonumber\\
T^{2}(s,t)&=&A(t,u,s)+A(u,s,t)\,,
\ea
and can be expanded in partial waves 
\be
T^{I}(s,t) = 32\pi\sum_{
  \ell=0}^{+\infty}(2\ell+1)P_{\ell}(\cos{\theta})t_{\ell}^{I}(s),
\ee
where $t$ and $u$ have been written as
$t=-\frac{1}{2}(s-4m^{2}_{\pi})(1-\cos{\theta})$,
 $u=-\frac{1}{2}(s-4m^{2}_{\pi})(1+\cos{\theta})$.
Near threshold the $t^I_{\ell}$ are further expanded in terms of
the threshold parameters
\be
t_{\ell}^{I}(s)=q^{2\ell}\left( a_{\ell}^{I}+ b_{\ell}^{I}q^{2}
+ c_{\ell}^{I}q^{4}+ d_{\ell}^{I}q^{6}+\mathcal{O}(q^{8})\right)\quad
 q^{2}=\frac{1}{4}(s-4m^{2}_{\pi}) ,
\ee
where $a_{\ell}^{I}, b_{ \ell}^{I}\dots$ are the 
scattering lengths, slopes,$\dots$ and $q$ is the magnitude of the pion
three momenta in the center of mass frame. 
We studied only those observables where a dependence on the $C_{i}$s shows
up.
Using $s+t+u = 4 m_\pi^2$ we can write the amplitude to order $p^6$ as
\begin{eqnarray}
\label{Apipi}
A(s,t,u)=b_{1}+b_{2}s+b_{3}s^{2}+b_{4}(t-u)^{2}+b_{5}s^{3}+b_{6}s(t-u)^{2}
+\textrm{non polynomial part}
\end{eqnarray}
The tree level Feynman diagrams give polynomial contributions to
$A(s,t,u)$ which must be expressible in terms of $b_1,\dots,b_6$.

The threshold parameters $a^0_0$, $b^0_0$, $c^0_0$, $d^0_0$,
$a^2_0$, $b^2_0$, $c^2_0$, $d^2_0$, $a^1_1$, $b^1_1$, $c^1_1$,
$a^0_2$, $b^0_2$, $a^2_2$, $b^2_2$, $a^1_3$
are all those that can receive contributions from tree level LECs up to
order $p^6$, but results \cite{CGL} have only been presented for
 $a^0_0$, $b^0_0$, 
$a^2_0$, $b^2_0$, $a^1_1$, $b^1_1$,
$a^0_2$, $b^0_2$, $a^2_2$, $b^2_2$ and $a^1_3$.
At present we thus can only use those 11 to test ChPT. 
We do not consider $b^1_3$ for which numerical results are also given
in \cite{CGL} since it does not depend at tree level on any LECs
to order $p^6$.
For those 11 we
obtain the following five relations:
\ba
\label{pipi1}
 \treeC{5b^2_0- 2b^0_0 
  - 27 a^1_1 - 15a^2_0 + 6a^0_0}&=& -  18\treeC{b^1_1}
,\\
\label{pipi2}
\treeC{3a^1_1+b^2_0} &=&
 20\treeC{b^2_2 - b^0_2- a^2_2+a^{0}_2},\\
\label{pipi3}
\treeC{b^0_0+5b^2_0 +9a^1_1}
& = &
90\treeC{a^0_2-b^0_2}
,\\
\label{pipi4}
       \treeC{3b^{1}_1 + 25a^{2}_2} &=& 10\treeC{a^0_2},\\
\label{pipi5}
  \treeC{- 5b^2_2 + 2b^0_2} &=& 21\treeC{a^1_3},
\ea
All quantities are expressed in units of $m_{\pi^+}^2$.
In fact, since these relations hold for every
contribution to the polynomial part, they are valid for the NLO tree level
contribution as well and for two- and three-flavour ChPT.
Therefore they do not get contributions from the $L_i$s at NLO, but
only at NNLO via the non polynomial part of Eq.~(\ref{Apipi}).

The first three involve quantities that already have tree level contributions
at lowest order, the fourth starts with tree level at NLO
and the last only has tree level
contributions starting at NNLO. The terms in the first three are arranged such
that the quantities starting at lowest order are all on the left-hand-side.

Let us now look at the numerical results. As experimental input we use
 the Roy equation
analysis together with input from ChPT and the pion scalar form-factor
 done in \cite{CGL}.
In Tab.~\ref{tabpipi} we quote the left-hand-side (LHS) and right-hand-side
 (RHS) of each of
the relations with the threshold parameters as quoted in \cite{CGL}. 
We have added the errors
for the several quantities quadratically which probably results in an
 underestimate of the
error. The results are quoted in the second column of Tab.~\ref{tabpipi}.
The next columns give the contribution from pure one-loop at NLO,
 the tree level NLO contribution
at one-loop using the fitted values of fit 10 in \cite{ABT4},
 the pure two-loop contribution,
and the $L_i$ dependent part at NNLO (called NNLO 1-loop) using
 again fit 10 of \cite{ABT4}.
Of these the tree level NLO contribution
must satisfy the relations, the others need not. 
The numerical results have been calculated using the formulas of \cite{BDT1}.
The column labeled remainder is the result of \cite{CGL} minus
 the three-flavour ChPT
prediction. This is thus the contribution of the NNLO LECs and from
higher orders.

The theoretical errors are more difficult to estimate.
The error shown in the sixth column in brackets in
Tab.~\ref{tabpipi} is obtained by varying all the $L_i^r$ around the central
values of fit 10 of \cite{ABT4} exploring the region with
$\chi^2/dof\approx 1$ using the full covariance matrix as obtained
for that fit by the authors of \cite{ABT4}. 
The error is then estimated as the
maximum deviation observed. The error for the $L_i^r$
contribution at NLO is not shown since it drops out of the relations.

As we see, the first three relations are very well satisfied. 
The last two work at a level around two sigma. 
Uncertainties on the theoretical results
are mostly on the last quoted digit, no uncertainty due to fit 10 is included.
Note that the $\pi\pi$ threshold parameters were not used as input in fit 10.

\begin{table}
\centering
\begin{tabular}{|c|c|r|r|r|r|c|}
\hline
                  & \cite{CGL}            & NLO    & NLO    &NNLO    &NNLO    & remainder\\
                  &                       & 1-loop &LECs    &2-loop  &1-loop  &          \\
\hline
LHS (\ref{pipi1})       & $ 0.009\pm0.039$&$ 0.054$&$-0.044$&$-0.041$&$-0.002(3) $&$ 0.041\pm0.039$\\
RHS (\ref{pipi1})       & $-0.102\pm0.002$&$-0.009$&$-0.044$&$-0.060$&$-0.008(6) $&$ 0.018\pm0.002$\\
10 LHS (\ref{pipi2})    & $ 0.334\pm0.019$&$ 0.209$&$ 0.097$&$ 0.103$&$ 0.029(11)$&$-0.105\pm0.019$\\
10 RHS (\ref{pipi2})    & $ 0.322\pm0.008$&$ 0.177$&$ 0.097$&$ 0.120$&$ 0.034(13)$&$-0.107\pm0.008$\\
LHS (\ref{pipi3})       & $ 0.216\pm0.010$&$ 0.166$&$ 0.029$&$ 0.053$&$ 0.016(6)$&$-0.047\pm0.010$\\
RHS (\ref{pipi3})       & $ 0.189\pm0.003$&$ 0.145$&$ 0.029$&$ 0.049$&$ 0.020(7)$&$-0.054\pm0.003$\\
10 LHS (\ref{pipi4})    & $ 0.213\pm0.005$&$ 0.137$&$ 0.032$&$ 0.053$&$ 0.035(12)$&$-0.043\pm0.005$\\
10 RHS (\ref{pipi4})    & $ 0.175\pm0.003$&$ 0.121$&$ 0.032$&$ 0.050$&$ 0.029(10)$&$-0.057\pm0.003$\\
$10^3$ LHS (\ref{pipi5})& $ 0.92\pm0.07  $&$ 0.36$ &$ 0.00$ &$ 0.56$ &$-0.01(13)$ &$0.00\pm0.07 $\\
$10^3$ RHS (\ref{pipi5})& $ 1.18\pm0.04  $&$ 0.42$ &$ 0.00$ &$ 0.57$ &$ 0.03(13)$ &$0.15\pm0.04 $\\
\hline
\end{tabular}
\caption{\label{tabpipi} The relations found in the $\pi\pi$-scattering.
The lowest order contribution is always zero by construction.
The NLO LEC part satisfies the relation. Notice the extra factors of
ten for some of them. All quantities are in the units of powers of $m_{\pi^+}$.
See text for a longer discussion.
}
\end{table}

We can also check how the two-flavour predictions hold up.
Here the expansion parameter is different. The corrections are in powers
of $m_\pi^2$ rather than in powers of  $m_K^2$. The expansion should thus
converge better and the conclusion was drawn in \cite{CGL} that two-flavour
ChPT works for $\pi\pi$-scattering at threshold (and even better
where they performed their subtractions). We do not use the numbers quoted
in \cite{BCEGS1,ABT3} since the LECs used there have been superceded
by those of \cite{CGL} and \cite{ABT4} respectively.
Testing our relations for
two-flavour ChPT thus gives a good indication of the best results we
can expect for the three-flavour case.
We use the threshold parameters as
quoted in \cite{CGL} for their best fit of the NLO LECs and using the
formulas of \cite{BCEGS1}. The result is shown in Tab.~\ref{tabpipi2}.
We see the same pattern as for the three flavour case.
The first three relations are very well satisfied while the last two
are somewhat worse but here below two sigma.
\begin{table}
\centering
\begin{tabular}{|c|c|c|c|}
\hline
                  & \cite{CGL}            & two-flavour & remainder\\
                  &                       & \cite{CGL} &          \\
\hline
LHS (\ref{pipi1})       & $ 0.009\pm0.039$&$-0.003$&$ 0.007\pm0.039$\\
RHS (\ref{pipi1})       & $-0.102\pm0.002$&$-0.097$&$-0.005\pm0.002$\\
10 LHS (\ref{pipi2})    & $ 0.334\pm0.019$&$0.332$&$ 0.002\pm0.019$\\
10 RHS (\ref{pipi2})    & $ 0.322\pm0.008$&$0.318$&$ 0.004\pm0.075$\\
LHS (\ref{pipi3})       & $ 0.216\pm0.010$&$0.206$&$ 0.010\pm0.010$\\
RHS (\ref{pipi3})       & $ 0.189\pm0.003$&$0.189$&$ 0.000\pm0.003$\\
10 LHS (\ref{pipi4})    & $ 0.213\pm0.005$&$0.204$&$ 0.009\pm0.005$\\
10 RHS (\ref{pipi4})    & $ 0.175\pm0.003$&$0.176$&$-0.001\pm0.003$\\
$10^3$ LHS (\ref{pipi5})& $ 0.92\pm0.07  $&$1.00$ &$-0.08 \pm0.07 $\\
$10^3$ RHS (\ref{pipi5})& $ 1.18\pm0.04  $&$1.15$ &$ 0.04 \pm0.04 $\\
\hline
\end{tabular}
\caption{\label{tabpipi2} The relations found in the $\pi\pi$-scattering
  evaluated in two-flavour ChPT. In the second column we have used the
 NNLO results
  quoted in \cite{CGL}. Notice the extra factors of
ten for some of them. See text for a longer discussion.
}
\end{table}

\begin{figure}
\begin{center}
\includegraphics[angle=270,width=12cm]{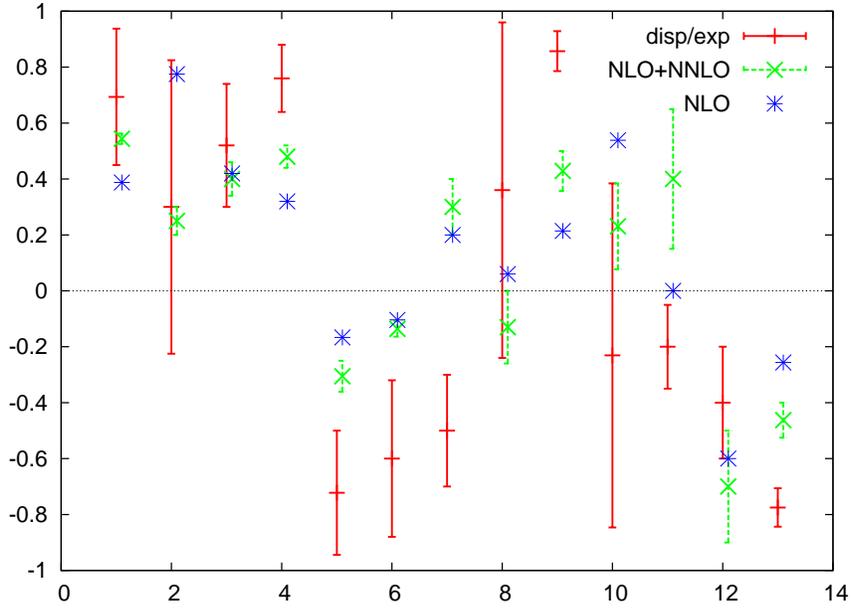}
\end{center}
\caption{\label{figfullrelations} The relations with the dispersive/experimental
results shown as full lines with errors, the NLO result as stars
and the sum of NLO+NNLO as crosses with the errors indicated as dashed lines.
The scale is arbitrary. The relations appear in the order
given in Tab.~\ref{tabfullrelations}. 1-5 $\pi\pi$, 6-10 $\pi K$,
11-12 $\pi\pi$ vs $\pi K$ and 13 $K_{\ell4}$ vs $\pi K$.}
\end{figure}

An alternative way to look at the results is to directly test the relations.
In the previous tables we have presented results separately for the LHS
and RHS in order to show how well the combinations of the NNLO LECs
would be the same if determined in the two different ways.
We can also instead show LHS minus RHS for our relations which
directly tests the loop content of ChPT. For the $\pi\pi$ and $\pi K$
case this is equivalent to comparing the exact results for the dispersive
part with the ChPT result for the dispersive part since the subtraction
constants used in \cite{CGL} drop out in the relations we
consider\footnote{We thank
the referee for pointing this out.}.
This is shown in Tab.~\ref{tabfullrelations}.
\begin{table}
\begin{center}
\begin{tabular}{|c|c|c|c|c|}
\hline
                           & disp/exp        & NLO     & NLO+NNLO & NLO+NNLO(2)\\
\hline
LR (\ref{pipi1})           & $0.111\pm0.039$ & $0.062$ & 0.087(3) & 0.094 \\
10 LR (\ref{pipi2})       & $0.012\pm0.021$ & $0.031$ & 0.010(2) & 0.014 \\
LR (\ref{pipi3})          & $0.026\pm0.011$ & $0.021$ & 0.020(3) & 0.017 \\
10 LR (\ref{pipi4})       & $0.038\pm0.006$ & $0.016$ & 0.024(2) & 0.028 \\
$10^3$ LR (\ref{pipi5})   & $-0.26\pm0.08$  & $-0.06$ &$-0.11(2)$&$-0.14$\\
\hline 
LR (\ref{piK1})           & $-1.5\pm0.7$    & $-0.26$ & $-0.34(7)$& - \\
10 LR (\ref{piK2})        & $-0.05\pm0.02$  & $0.02$  & $0.03(5)$ & - \\
100 LR (\ref{piK3})       & $0.36\pm0.60$   & $0.06$  & $-0.13(13)$& - \\
100 LR (\ref{piK4})       & $0.12\pm0.01$   & $0.03$  & $0.06(1)$& - \\
$10^3$ LR (\ref{piK5})    & $-0.03\pm0.08$  & $0.07$  & $0.03(2)$& - \\
\hline
$10^3$ LR (\ref{pipipiK1})&$-0.04\pm0.03$   & $0.00$  & $0.08(5)$ & - \\
10 LR (\ref{pipipiK2})    &$-0.04\pm0.02$   & $-0.06$ & $-0.07(2)$& - \\
\hline
LR (\ref{Kl4rel})         &$-1.24\pm0.11$   & $-0.41$ & $-0.74(10)$& - \\
\hline
\end{tabular}
\end{center}
\caption{\label{tabfullrelations} Tests of the relations
as seen as a test of the loop contributions. 
disp/exp are the dispersive and experimental inputs used
as described in the text.
LR stands for
LHS-RHS. All quantities are in units of $m_{\pi^+}$.
Results are shown for the relations for $\pi\pi$, $\pi K$, $\pi\pi$ vs $\pi K$
and $K_{\ell4}$ vs $\pi K$.}
\end{table}
We see here also good agreement for the first three and about two sigma
for the last two relations.
The results given in Tab.~\ref{tabfullrelations} are depicted graphically
in Fig.~\ref{figfullrelations}. Keep in mind here that the errors
for the dispersive result might be underestimated since we
combined them quadratically.

\section{$\pi K$ scattering}
\label{piK}

The $\pi K$ scattering amplitude has amplitudes $T^I(s,t,u)$
in the isospin channels $I= 1/2,3/2$.
 As for $\pi \pi$ scattering, it is possible to define scattering lengths
 $a_{\ell}^{I}$, $b_{\ell}^{I}$. So we introduce the partial wave expansion of
 the isospin amplitudes
\be
T^{I}(s,t,u) = 16\pi
\sum_{\ell=0}^{+\infty}(2\ell+1)P_{\ell}(\cos{\theta})t_{\ell}^{I}(s),
\ee
and we expand the $t_{\ell}^{I}(s)$ near threshold:
\be
\label{tilpiK}
t_{\ell}^{I}(s)=\frac{1}{2}\sqrt{s}q_{\pi K}^{2\ell}
\left(a_{\ell}^{I}+b^{I}_{\ell}q^{2}_{\pi K}+c^{I}_{\ell}q^{4}_{\pi K}
+\mathcal{O}(q^{6}_{\pi K})\right),
\ee
where
\be
\label{qpiK}
q_{\pi K}^{2}=\frac{s}{4}\left(1-\frac{(m_{K}+m_{\pi})^{2}}{s}\right)
          \left(1-\frac{(m_{K}-m_{\pi})^{2}}{s}\right)\,,
\ee
is the magnitude of the three-momentum in the center of mass system.
The Mandelstam variables are in terms of the scattering angle given by
\be
t=-2q^{2}_{\pi K}(1-\cos{\theta}),\quad u=-s-t+2m^{2}_{K}+2m^{2}_{\pi}\,.
\ee
Again we studied only those observables where a
dependence on the $C_{i}$s shows up.
 
It is also customary to introduce the crossing symmetric and antisymmetric
amplitudes $T^{\pm}(s,t,u)$ which can be expanded around
$t=0$, $s=u$ using $\nu= (s-u)/(4m_{K})$, called the subthreshold expansion:
\ba
\label{subthrexp}
T^{+}(s,t,u)&=&\sum_{i,j=0}^{\infty}c_{ij}^{+}t^{i}\nu^{2j}, \qquad
T^{-}(s,t,u)=\sum_{i,j=0}^{\infty}c_{ij}^{-}t^{i}\nu^{2j+1}.
\ea
There are ten subthreshold parameters that have tree level contributions
 from the NNLO LECs.
In $c_{01}^{-}$ and $c_{20}^{-}$ the same combination
$-C_{1}+2C_{3}+2C_{4}$ appears~\cite{BDT1}, thus
\be
\label{piKrel}
16\rho^{2}\treeC{c_{20}^{-}}=3\treeC{c_{01}^{-}}\,.
\ee
Eq.~(\ref{piKrel}) leads to one relation between the subthreshold parameters.

If we look at the $a^I_\ell$ and $b^I_\ell$ that get contributions
from the NNLO LECs there are 14 such. 7 for each isospin channel.
The isospin odd channel only involves $T^{-}$:
\be
T^{1/2}(s,t,u)-T^{3/2}(s,t,u) = 3 T^{-}(s,t,u)\,.
\ee
This combination has only three subtreshold parameters that get independent
 contributions from
the NNLO LECs. So for 7 differences of $a^I_\ell$ and $b^I_\ell$ and three
 parameters
we expect four relations. 
The threshold parameters are expressed in units of $m_{\pi^+}$ and we use
the symbol $\rho=m_K/m_\pi$. We use the notation
 $a^-_\ell = a_\ell^{1/2}-a^{3/2}_\ell$
and $b^-_\ell = b_\ell^{1/2}-b^{3/2}_\ell$
\ba
\label{piKrel1}
70\rho^3\left(\rho+1\right)^2 \treeC{a_3^-}
 &=&
-\left(\rho^2+\rho+1\right)\treeC{a_0^-}
+2\rho^2\treeC{b_0^-}+6\rho^2\treeC{a_1^-}\,,
\\
\label{piKrel2}
140\rho^3\left(\rho^2+1\right) \treeC{a_3^-}
 &=&
\left(\rho^2+1\right)\treeC{a_0^-}
+6\left(-\rho^2+\rho-1\right)\rho \treeC{a_1^-}+12\rho^3\treeC{b_1^-}\,,
\\
\label{piK2}
 5\left(\rho^2+1\right)\treeC{a_2^-}
 &=&
\treeC{a_1^-}
+2\rho\treeC{b_1^-}\,,
\\
\label{piK4}
 7\left(\rho^2+1\right)\treeC{a_3^-}
 &=&
\treeC{a_2^-}
+2\rho\treeC{b_2^-}\,.
\ea
We can eliminate $\treeC{a_3^-}$ from (\ref{piKrel1}) and (\ref{piKrel2})
to obtain a relation
involving only $\ell=0,1$ threshold parameters:
\ba
\label{piK1}
\left(\rho^4+3\rho^3+3\rho+1\right)\treeC{a_1^-}&=&
2\rho^2\left(\rho+1\right)^2\treeC{b_1^-}
-\frac{2}{3}\rho\left(\rho^2+1\right)\treeC{b_0^-}
\nonumber\\&&
+\frac{1}{2\rho}\left(\rho^2+\frac{4}{3}\rho+1\right)\left(\rho^2+1\right)
\treeC{a_0^-}\,.
\ea
We prefer to express the other relation in one involving $b_2^-$
\ba
\label{piK3}
5\left(\rho+1\right)^2\treeC{b_2^-}
 &=&\frac{\left(\rho-1\right)^2}{\rho^2}\treeC{a_1^-}
-\frac{\rho^4+\frac{2}{3}\rho^2+1}{4\rho^4}\treeC{a_0^-}
+\frac{\rho^2-\frac{2}{3}\rho+1}{2\rho^2}\treeC{b_0^-}\,.
\ea

The combination that involves only $T^+$ is
\be
T^{1/2}(s,t,u)+2T^{3/2}(s,t,u) = 3 T^{+}(s,t,u)\,.
\ee
This brings in 7 more threshold parameters, but there are 6 fully
 independent subtreshold
parameters so we expect only one more relation. Using
the notation $a^+_\ell = a_\ell^{1/2}+2a^{3/2}_\ell$
and $b^+_\ell = b_\ell^{1/2}+2b^{3/2}_\ell$, we find:
\ba
\label{piK5}
7\treeC{a_3^+} &=&\frac{1}{2\rho}\treeC{a_2^+}
-\treeC{b_2^+}
+\frac{1}{5\rho}\treeC{b_1^+}
-\frac{1}{60\rho^3}\treeC{a_0^+}
-\frac{1}{30\rho^2}\treeC{b_0^+}\,.
\ea

These relations hold for all tree-level contributions up to 
NNLO\footnote{This was written wrong
in the preliminary report \cite{EFT09ilaria}.}.
In particular, the lowest order contributions satisfy them.

Note that because of the nonlinearity in $s$ present in (\ref{qpiK})
the higher order threshold parameters are already nonzero at lowest order.
This makes fitting the threshold-expansion numerically more unstable since we
need to use a fitting polynomial to higher order in $q^2_{\pi K}$ compared
to what was needed for the $\pi\pi$ case.

\begin{table}
\centering
\begin{tabular}{|c|c|r|r|r|r|c|}
\hline
                  & \cite{BDM}         & NLO   & NLO  &NNLO    &NNLO    & remainder\\
                  &                    & 1-loop&LECs  &2-loop  &1-loop  &          \\
\hline
LHS (\ref{piK1})       & $ 5.4\pm0.3  $&$0.16$&$ 0.97$&$ 0.77$&$-0.11(11)$&$ 0.6\pm0.3$\\
RHS (\ref{piK1})       & $ 6.9\pm0.6  $&$0.42$&$ 0.97$&$ 0.77$&$-0.03(7)$&$ 1.8\pm0.6$\\
10 LHS (\ref{piK2})    & $ 0.32\pm0.01$&$0.03$&$ 0.12$&$ 0.11$&$ 0.00(2)$&$ 0.07\pm0.01$\\
10 RHS (\ref{piK2})    & $ 0.37\pm0.01$&$0.02$&$ 0.12$&$ 0.10$&$-0.01(2)$&$ 0.14\pm0.01$\\
100 LHS (\ref{piK3})   & $-0.49\pm0.02$&$0.08$&$-0.25$&$-0.17$&$ 0.05(3)$&$-0.21\pm0.02$\\
100 RHS (\ref{piK3})   & $-0.85\pm0.60$&$0.03$&$-0.25$&$ 0.11$&$-0.03(13)$&$-0.71\pm0.60$\\
100 LHS (\ref{piK4})   & $ 0.13\pm0.01$&$0.04$&$ 0.00$&$ 0.01$&$ 0.03(1)$&$ 0.05\pm0.01$\\
100 RHS (\ref{piK4})   & $ 0.01\pm0.01$&$0.01$&$ 0.00$&$ 0.00$&$ 0.00(1)$&$-0.01\pm0.01$\\
$10^3$ LHS (\ref{piK5})& $ 0.29\pm0.03$&$0.09$&$ 0.00$&$ 0.06$&$ 0.01(2)$&$ 0.13\pm0.03$\\
$10^3$ RHS (\ref{piK5})& $ 0.31\pm0.07$&$0.03$&$ 0.00$&$ 0.06$&$ 0.05(3)$&$ 0.17\pm0.07$\\
\hline
\end{tabular}
\caption{\label{tabpiK} The relations found in the $\pi K$-scattering.
The tree level contribution to the LHS and RHS
 of relation 1
is 3.01 and vanishes for the others.
The NLO LECs part satisfies the relation. Notice the extra factors of
ten for some of them. 
See text for a longer discussion.
All quantities are in the units of powers of $m_{\pi^+}$.
}
\end{table}
The column labeled \cite{BDM} uses the results of the Roy-Steiner
analysis of \cite{BDM} of $\pi K$ scattering. We have combined 
errors
 quadratically
which due to the presence of correlations can lead to a serious underestimate
of the errors on the combinations.

 The numerical results
for the theory are calculated with the formulas of \cite{BDT2}
where the NLO LECs we use are those of fit 10 of \cite{ABT4}.
The columns in Tab.~\ref{tabpiK} have the same meaning as in
 Tab.~\ref{tabpipi} and the errors on the ChPT part have been evaluated
as discussed for the $\pi\pi$ case.
The first relation is reasonably well satisfied, somewhat below two sigma.
 The second
relation has a large discrepancy in view of the experimental error but
if we assume
a theory error of about half the NNLO contribution it seems reasonable
given 
 The third relation
is well satisfied but the RHS has a rather large experimental error.
The fourth relation does not work well, mainly due to the fact that we seem to
underestimate the value for $a^-_3$. The last relation again works
 reasonably well. The same relations but now LHS-RHS
are shown in Tab.~\ref{tabfullrelations} and depicted graphically
in Fig.~\ref{figfullrelations}. The conclusions are the same.

\section{$\pi\pi$ and $\pi K$ scattering}
\label{pipipiK}

If we consider the $\pi\pi$ and $\pi K$ system
 together we get two more relations due to the identities
\ba
\treeC{b_{5}}&=&\treeC{c^{+}_{30}}+\frac{3}{\rho}\treeC{c^{-}_{20}}\,,
\qquad
\treeC{b_{6}}=\frac{1}{4\rho}\treeC{c^{-}_{20}}
   +\frac{1}{16\rho^2}\treeC{c^{+}_{11}},
\ea
where $c^{-}_{ij}$ ($c^{+}_{ij}$) are  expressed
in units of $m^{2i+2j+1}_{\pi}$($m^{2i+2j}_{\pi}$). 
We can express these relations in terms of the threshold parameters:
\ba
\label{pipipiK1}
6 \treeC{a^1_3}&=&\left(1+\rho\right)\treeC{a^+_3+3 a^-_3}\,,
\\
\label{pipipiK2}
3\left[\left(1+\rho\right)^2 \treeC{b^2_2}
 + 7\left(1-\rho\right)^2 \treeC{a^1_3} \right]&=&
\left(1+\rho\right)\left[7 \left(1-4\rho+\rho^2\right)\treeC{a^-_3}
+\treeC{a^+_2+2\rho b^+_2}\right]\,.
\nonumber\\
\ea
Here all the quantities are expressed in powers of $m_{\pi^+}$.

The numerical results are quoted in Tab.~\ref{tabpipipiK}. The first
 relation does not work
but the second is well satisfied. If we look in the numerical results
 we see that $a^-_3$
plays a minor role in the RHS of the second relation but is important
 in the first, so this
could be the same problem that appeared for relation (\ref{piK4}).
 The same relations but now LHS-RHS
are shown in Tab.~\ref{tabfullrelations} and depicted graphically
in Fig.~\ref{figfullrelations}. The conclusions are the same.
A related analysis can be found in \cite{KM}.

\begin{table}
\centering
\begin{tabular}{|c|c|r|r|r|r|c|}
\hline
                  & \cite{CGL},\cite{BDM}  & NLO   & NLO  &NNLO    &NNLO    & remainder\\
           & \cite{Pislak2},\cite{NA48kl4} & 1-loop&LECs  &2-loop  &1-loop  &          \\
\hline
$10^3$ LHS (\ref{pipipiK1})& $0.34\pm0.01$&$ 0.12$&$0.00$&$ 0.16$&$ 0.00(4)$&$ 0.05\pm0.01$\\
$10^3$ RHS (\ref{pipipiK1})& $0.38\pm0.03$&$ 0.12$&$0.00$&$ 0.05$&$ 0.04(2)$&$ 0.16\pm0.03$\\
10 LHS (\ref{pipipiK2})   & $-0.13\pm0.01$&$-0.12$&$0.00$&$-0.05$&$ 0.02(2)$&$ 0.01\pm0.01$\\
10 RHS (\ref{pipipiK2})   & $-0.09\pm0.02$&$-0.05$&$0.00$&$-0.02$&$-0.01(1)$&$-0.01\pm0.02$\\
\hline
LHS (\ref{Kl4rel}) & $-0.73\pm0.10$ & $-0.23$ & $0.00$ & $-0.15$ & $-0.05(6)$ & $-0.29\pm0.10$\\ 
RHS (\ref{Kl4rel}) &  $0.50\pm0.07$ &  $0.19$ & $0.00$ &  $0.10$ &  $0.03(4)$ & $0.18\pm0.07$\\
\hline
\end{tabular}
\caption{\label{tabpipipiK} The relations found between $\pi\pi$ and $\pi
  K$-scattering lengths and between the curvature in $F$ in $K_{\ell4}$
 and $\pi K$ scattering.
  See text for a longer discussion.
  All quantities are in units of powers of $m_{\pi^+}$.
}
\end{table}

\section{$K_{\ell4}$}
\label{Kl4}

The decay $K^+(p)\to \pi^+(p_1)\pi^-(p_2) e^+(p_\ell)\nu(p_\nu)$ is given 
by the amplitude~\cite{Bijnens:1994me}
\be
      T = \frac{G_F}{\sqrt{2}} V^\star_{us} \bar{u} (p_\nu) \gamma_\mu
      (1-\gamma_5) v(p_\ell) (V^\mu - A^\mu)
      \label{k11}
\ee
where $V^\mu$ and $A^\mu$ are parametrized in terms of four form factors: $F$,
$G$, $H$ and $R$  (but the $R$-form factor is negligible in decays with
an electron in the final state).
Using partial wave expansion and neglecting $d$ wave terms one
obtains~\cite{Amoros:1999mg}:
\ba
\label{Kl4exp}
F&=&f_{s}+f'_{s}q^{2}+f''_{s}q^{4}+f'_{e}s_{e}/4m^{2}_{\pi}
+f_{t}\sigma_\pi X\cos\theta+\ldots
\,,
\nonumber\\
G_{p}&=&g_{p}+g'_{p}q^{2}+g''_{g}q^{4}+g'_{e}s_{e}/4m^{2}_{\pi}
+g_{t}\sigma_\pi X\cos\theta+\ldots
\ea
Here
$s_{\pi}(s_{e})$ is the invariant mass of dipion (dilepton) system, and
$ q^{2}=s_{\pi}/(4m^{2}_{\pi})-1$.
$\theta$ is the angle of the pion in their restframe
w.r.t. the kaon momentum and $t-u=-2\sigma_\pi X\cos\theta$.  
We found one relation between the quantities defined in (\ref{Kl4exp}) and
$\pi K$ scattering:
\be
\label{Kl4r}
\sqrt{2}\treeC{f''_{s}}= 64 \rho F_\pi\treeC{c^{+}_{30}}\,.
\ee
This translates into a relation between $\pi K$ threshold parameters
and $f''_{s}$ which, with all quantities expressed in units of $m_{\pi^+}$,
reads:
\be
\label{Kl4rel}
\sqrt{2}\treeC{f''_{s}}= 32\pi\frac{\rho}{1+\rho} F_\pi
\left[\frac{35}{6}\left(2+\rho+2\rho^2\right)\treeC{a^+_3}
-\frac{5}{4}\treeC{a^+_2+2\rho b^+_2}\right]\,.
\ee
There is no more relation involving the quantities discussed so far, $\pi\pi$
 and $\pi K$ scattering, and $K_{\ell4}$.

Numerical results for (\ref{Kl4rel}) are shown in Tab.~\ref{tabpipipiK}.
The experimental results is taken from \cite{NA48kl4} for $f''_s/f_s$ and from
 \cite{Pislak2}
for $f_s$. This should be an acceptable combination since the central value
 for $f'_s/f_s$ and
$f''_s/f_s$ from \cite{Pislak2} are within 10\% of those of \cite{NA48kl4}.
The theoretical results are using the formulas of \cite{ABT3} and fit 10
 of \cite{ABT4}.
This relation has problems. The sign is even different on both sides.
 In both cases we also
see that the ChPT series has a large NNLO contribution.
For completeness, LHS-RHS is given in Tab.~\ref{tabfullrelations} and
Fig.~\ref{figfullrelations}.

There have been indications from dispersive methods that ChPT might
underestimate the curvature $f''_s$. Dispersion relations
were used in \cite{BCG} for $K_{\ell4}$.
If one looks at Fig.~7 in \cite{ABT4}, one can see that the dispersive result
of \cite{BCG} has a larger curvature than the two-loop result.
For this reason, we do not consider this discrepancy a major problem for ChPT.

\section{$\eta\to3\pi$}
\label{e3pi}

The amplitude for the decay
 $\eta(p_\eta)\to\pi^+(p_+)\pi^-(p_-)\pi^0(p_0)$
can be written as
\be
A(\eta\to\pi^+\pi^-\pi^0) = \sin\epsilon M(s,t,u)\,.
\ee
Here we used the Mandelstam variables
\ba
s &=& (p_+ +p_-)^2  = (p_\eta - p_0)^2   \,,
\nonumber\\
t &=& (p_+ +p_0)^2   = (p_\eta -p_-)^2     \,,
\nonumber\\
u &=& (p_- +p_0)^2   = (p_\eta -p_+)^2      ,.
\ea
which are linearly dependent
$
s+t+u = m_{\pi^{o}}^2 + m_{\pi^{-}}^2 + m_{\pi^{+}}^2 + m_{\eta}^2
\equiv 3 s_0\,.  
$.
G-parity requires the amplitude to vanish at the 
limit $m_u = m_d$ and therefore it must inevitably
be accompanied by an overall factor of $m_u-m_d$
which we have chosen to be in the form of
$\sin(\epsilon)\approx (\sqrt{3}/4)(m_d-m_u)/(m_s-\hat m)$.
Since the amplitude is invariant under charge 
conjugation we have 
$
M(s,t,u) = M(s,u,t).
$
Similar to the $\pi\pi$ scattering, we can write the amplitude as
\be
M(s,t,u) = \eta_1 +\eta_2 s +\eta_3 s^2 +\eta_4 (t-u)^2+\eta_5 s^3 
+ \eta_6 s (t-u)^2
+ \textrm{non~polynomial~part}
\ee
to NNLO in ChPT. Using the results of \cite{BG2} we then obtain two relations
\ba
\treeC{\eta_5}&=&3\treeC{\eta_6}\,,
\\
\treeC{\eta_5} &=& -768 \rho^3 \treeC{c^-_{01}}
 = -\pi\left(1+\rho\right)\frac{35}{2}\treeC{a^-_3}\,.
\ea
Since $\eta_5$ is not unambiguously determined from the measured Dalitz-plot
 parameters
and $\eta_6$ is not measured at all we do not present numerical results for 
this.
The overall factor $\sin\epsilon$ itself is part of the uncertainty involved.
Unfortunately, no relations are present for $\eta_1,\ldots,\eta_4$ which would
 have helped
in the numerical prediction for $\eta\to3\pi$ using the results of \cite{BG2}.

\section{Scalar formfactors}
\label{scalarFF}

The scalar form factors for the pions and the kaons are defined as
\begin{equation}
F_{ij}^{M_{1}M_{2}}(t)=\langle M_{2}(p)|\bar{q}_{i}q_{j}|M_{1}(q)\rangle,
\end{equation}
where $t=p-q$, $i,j=u, d, s$ are flavour indices and $M_{i}$ denotes a meson
state with the indicated momentum.
Due to isospin symmetry not all of them are independent, therefore we
consider only 
\begin{eqnarray}
 F_{S}^{\pi} &=& 2F^{\pi^0\pi^0}_{uu}\,\qquad\quad
F_{Ss}^{\pi}=F^{\pi^0\pi^0}_{ss}\,,\qquad
F_{Ss}^{K}=F_{ss}^{K^0K^0},
\nonumber\\
F_{S}^{K} &=& F^{K}_{Su}+F^{K}_{Sd}=F^{K^0K^0}_{uu}+F^{K^0K^0}_{dd}\,,\qquad
F_{S}^{\pi K}=F_{su}^{K^+\pi^0}.
\end{eqnarray}
Near $t=0$ these are expanded via
\be
F_S(t) = F_S(0)+F_S' t + F_S'' t^2+\ldots\,.
\ee
The NNLO ChPT calculation for these quantities was performed in \cite{BD} where
it was found that the curvatures $F_S''$ only depend on two of the NNLO LECs.
As a consequence there are four relations
\ba
\treeC{F_{S}^{\pi\prime\prime}} &=& 2\treeC{F_{Su}^{K\prime\prime}}
 = 2\treeC{F_{Ss}^{K\prime\prime}}\,,
\nonumber\\
\treeC{F_{Ss}^{\pi\prime\prime}} &=& \treeC{F_{Sd}^{K\prime\prime}}\,,
\nonumber\\
2\treeC{F_{S}^{K\pi\prime\prime}} &=& \treeC{F_{S}^{\pi\prime\prime}}
-2\treeC{F_{Ss}^{\pi\prime\prime}}\,.
\ea
There is also a relation involving the slopes
\be
\treeC{F_S^{\pi\prime}}-2 \treeC{F_{Ss}^{\pi\prime}}-2\treeC{F_{Sd}^{K\prime}}
+2\treeC{F_{Ss}^{K\prime}}-4\treeC{F_{S}^{K\pi\prime}}=0\,.
\ee
This is a consequence of the ``scalar Sirlin'' relation derived in general 
in \cite{BD}.

In addition to those already known we found a relation between the
 values at $t=0$
which with $\rho=m_K/m_\pi$ reads
\ba
2\rho^6 \treeC{F_S^\pi(0)} &=& 
\rho^4\left(2\rho^4-\rho^2-3\right)\treeC{F_{Ss}^\pi(0)}
+\left(3\rho^2-1\right) \treeC{F_{Su}^K(0)}
\nonumber\\&&
+\left(6\rho^4-3\rho^2-1\right)\treeC{F_{Sd}^K(0)}
+\left(\rho^2+1\right)\treeC{F^K_{Ss}(0)} \,.
\ea
The scalar formfactors had a significant dependence on what was
used as input for $L_4^r$ and $L_6^r$ \cite{BD}. The curvature
 relations were studied there
and found to sometimes work and sometimes not, see Tab.~2 and
 Sect.~5.5 in \cite{BD}.
We intend to come back to these relations when constraints
 on $L_4^r$ and $L_6^r$ have been
included in a general fit.

\section{Scalar formfactors, masses and decay constants}
\label{scalarFFmass}

The three masses $M^{2}_{\pi},\,M^{2}_{K},\,M^2_\eta$ and decay 
constants $F_\pi,\,F_K\,,F_\eta$
are not related, they all have a different dependence on the NNLO LECs.
We do find some relations however when we combine them with the scalar
 formfactors.
The two-loop calculations for masses and decay constants can be found in
\cite{ABT1,GK} for $\pi$ and $\eta$ and in \cite{ABT1} for the kaon.

There are two relations between the $F_{S}(0)$ and the
ChPT expansion of the masses $M^{2}_{\pi},\,M^{2}_{K}$:
\begin{eqnarray}
\label{scalmassrel}
2B_{0}\treeC{M^{2}_{\pi}}&=&\frac{1}{3}
      \left\{(2\rho^{2}-1)\treeC{F^{\pi}_{Ss}(0)}
                  +\treeC{F^{\pi}_{S}(0)}\right\}
\nonumber\\
2B_{0}\treeC{M^{2}_{K}}&=&\frac{1}{3}
      \left\{(2\rho^{2}-1)\treeC{F^{K}_{Ss}(0)}
               +\treeC{F^{K}_{S}(0)}\right\}.
\end{eqnarray}
Remember we express everything in units powers of $m_\pi$.
One could arrive to the same conclusion using the Feynman-Hellmann Theorem
(see e.g.~\cite{GL2} or~\cite{BD}) which implies
for $q=u,d,s$ and $M=\pi,K$
\begin{eqnarray}
F^{M}_{Sq}(t=0) &=&
  \langle M|\bar{u}u|M\rangle =
  \frac{\partial m^{2}_{M}}{\partial m_{q}}\,.
\end{eqnarray}
On the other hand the ChPT expansion leads to
\begin{eqnarray}
\treeC{M^{2}_{\pi}}=\sum_{i}C_{i}(m_{q})^{3}=f(m_{u},m_{d},m_{s}),
\end{eqnarray}
that is an homogeneous function of order three.
Thanks to the Euler Theorem, $\treeC{M^{2}_{\pi}}$ can be written in terms
of its derivatives $(f(\mathbf{x})=\frac{1}{3}\sum^{n}_{i=1}\frac{\partial
  f}{\partial x_{i}}x_{i}\quad\mathbf{x}\in \mathbb{R}^{n})$. 
These are exactly the relations in Eq.~(\ref{scalmassrel}).
Something similar holds for the $p^4$ expression but with a factor
$1/2$ instead of $1/3$.

There are two more relations if we also include the decay constants.
 The first one is
\ba
\label{scalarkpi}
\lefteqn{\left(\rho^2-1\right)^2\frac{B_0}{F_0}\treeC{F_K-F_\pi}
+\left(\rho^2+1\right)B_0\treeC{M_K^2-M_\pi^2}
= }
&&\nonumber\\&&
\left(\rho^4-1\right)\treeC{F_{S}^{K\pi}(0)}
+\left(\rho^2-1\right)^3 \treeC{F_S^{K\pi\prime}}
+\left(\rho^2-1\right)^3\left(\rho^2+1\right)\treeC{F_{S}^{K\pi\prime\prime}}\,.
\ea
This relation is the same as the one found in \cite{BT3} for the $K_{\ell3}$ 
scalar formfactor when one uses
\be
\label{quarkeom}
\partial^\mu \overline s \gamma_\mu u = \left(m_s-m_u\right)i\,\overline s u\,,
\ee
and rewrites the quark masses into the pion and kaon mass.
The second relation, in the simplest form we found, reads
\ba
\lefteqn{
\left(4 \rho^2\treeC{F_\pi}-4\treeC{F_K}\right)\frac{B_0}{F_0}
= 2 \rho^4\treeC{F_S^{\pi\prime}}
+\left(2\rho^6-\rho^4 +\rho^2 \right)\treeC{F_{Ss}^{\pi\prime}}
}
&&\nonumber\\&&
+\left(-2 \rho^4 +\rho^2 -1\right)\treeC{F_{Sd}^{K\prime}}
-\left(\rho^2+1\right)\treeC{F_{Su}^{K\prime}}
+\left(-3\rho^2+1\right)\treeC{F_{Ss}^{K\prime}}\,.
\ea

We have not presented numerical results for the relations in this section
 since the assumptions
underlying fit 10 of \cite{ABT4} were such that all the left hand sides
 evaluate to zero.
In addition the right hand sides tend to contain poorly known quantities.

\section{Vector formfactors}
\label{vector}

The vector formfactors have been discussed extensively in \cite{BT2}
 and \cite{BT3}.
There three relations between the curvatures and the the Sirlin relation
 between the
slopes \cite{PS1} were found. We also find the expected relationship between
the scalar formfactors and the scalar formfactor in $K_{\ell3}$ which
 followed from
(\ref{quarkeom}). The numerical results for the relation between the slopes
and curvatures were discussed extensively in
\cite{BT2,BT3} and found to work well.
So this sector had the expected total of 7 relations added to those 
discussed above.

\section{Scalar formfactors, $\pi\pi$ and $\pi K$ scattering}
\label{scalarFFpipipiK}

There are two more relations when we combine the scalar formfactors and
 the scattering
amplitudes for $\pi\pi$ and $\pi K$ scattering. All three quantities are
 needed.
These relations are rather complicated. The first relation
is:
\ba
&& \rho^4\treeC{105 a^1_3+15 b^2_2-3 a^1_1+3 b^2_0-8 a^2_0}
        +\left(1+\rho\right)
           \left(35 \rho^4\treeC{ a^-_3}-\frac{1}{3} \rho\treeC{ a^-_0}\right)
\nonumber\\&&
  +\frac{2}{\rho+1} \rho^3\treeC{ a^+_1}
  +\frac{10}{\rho+1} \rho^4 (2+\rho+2 \rho^2) \treeC{b^+_2+7 a^+_3}
  +\frac{2}{3}(\rho+1) \left(\rho^2+1\right)\treeC{a^+_0}
\nonumber\\&&
  -\frac{10}{\rho+1} \rho^3 \left(2+3 \rho+2 \rho^2\right) \treeC{a^+_2}
  -\frac{4}{\rho+1} \rho^3 (1+\rho+\rho^2) \treeC{b^+_1}
         =
\nonumber\\&&
\frac{\rho^2}{8\pi B_0 F_0^2\left(1-3 \rho^2\right)}
\Bigg[- (1-\rho^2) \treeC{F^K_{Ss}(0)}
           +2  (1-3 \rho^2+3 \rho^4) \treeC{F^K_{Sd}(0)}
\nonumber\\&&
           + (1-3 \rho^2+3 \rho^4-5 \rho^6+2 \rho^8) \treeC{F^\pi_{Ss}(0)}
           + \frac{1}{2}(1-5 \rho^2+8 \rho^4-4 \rho^6) 
                    \treeC{F^\pi_S(0)}\Bigg]\,.
\ea
The second relation involves even more quantities:
\ba
&& -(1-\rho^4)8\pi B_0 F_0^2\Bigg[
          \rho^2\treeC{ b^0_0-12 a^2_0+2 b^2_0+45 b^2_2-315 a^1_3}
\nonumber\\&&
\qquad\qquad\qquad      +210 \rho^2 (1+\rho) \treeC{a^-_3}
-2\frac{1+\rho}{\rho}\treeC{a^-_0}
 \Bigg]
\nonumber\\&&
        +8\pi B_0 F_0^2(\rho-1) \Bigg[120 \rho^4 \treeC{b^+_2+7 a^+_3}
            -6 \rho^2 \treeC{a^+_1+2 \rho b^+_1}
            +2 (1+\rho)^2 \treeC{2 a^+_0-15 \rho^2 a^+_2}
                   \Bigg]
\nonumber\\&&
=  (1-\rho^2)\Bigg[12 (1-\rho^4) (F^{K\prime\prime}_{Ss}-F^{K\prime\prime}_{Sd})
          +12 (1-\rho^2-2 \rho^4) F^{K\pi\prime}_{S}
         +6 \rho^2 (1+2 \rho^2) F^{K\prime}_{Ss}
\nonumber\\&&
 \qquad \qquad\qquad                 -12 \rho^4 F^{K\prime}_{Sd}
               +6 \rho^2 F^{K\prime}_{Su}-12 \rho^4 F^{\pi\prime}_{Ss}
        \Bigg]
\nonumber\\&&
\quad        +(1+\rho^2) 12 F^{K\pi}_S(0)
        +12 \rho^2 F^K_{Su}(0)
       +3 (-1+3 \rho^2+6 \rho^4) F^K_{Sd}(0)
\nonumber\\&&
\quad         +(1-\rho^2) (2-\rho^2-8 \rho^4-8 \rho^6) F^\pi_{Ss}(0)
       -2 (1+2 \rho^2+2 \rho^4+4 \rho^6) F^\pi_{S}(0)\,. 
\ea

\section{A final relation: $K_{\ell4}$, $\pi K$ scattering and scalar formfactors}
\label{finalrelation}

The final relation we found is between $K_{\ell4}$, $\pi K$
scattering and the scalar formfactors. 
The version below is the simplest we found.
\ba
&&     (1-\rho^2)^2 (1+\rho^2) B_0\Bigg[
          12\sqrt{2}\frac{F_0}{\rho}
    \left(g_p-g^\prime_p+\left((1-\frac{1}{4} \rho^2\right)g^{\prime\prime}_p
          +\frac{1}{2}\rho^2 f_t\right)
\nonumber\\&&
\qquad\qquad\qquad\qquad      -16 \pi\frac{1+\rho}{\rho} a^-_0
      +70 \pi   (1+\rho) (20 \rho^2+\rho^4) a^-_3
  \Bigg]
\nonumber\\&&
=      12 (1+\rho^2) \rho^4 F^{K\pi}_S(0)
      +12 \rho^6 F^K_{Su}(0)+24 \rho^8 F^K_{Sd}(0)
\nonumber\\&&
\quad      +2 \rho^4 (1-\rho^4) (1-4 \rho^4) F^\pi_{Ss}(0)
      -2 \rho^4 (1+2 \rho^2+2 \rho^4+4 \rho^6) F^\pi_S(0)\,.
\ea

\section{Conclusions}
\label{conclusions}

We have performed a systematic search for combinations that allow a test of
ChPT at NNLO
order. We have therefore looked at the three masses and three decay constants,
11 $\pi\pi$ threshold parameters, 14 $\pi K$ threshold parameters,
 6 $\eta\to3\pi$
parameters, 10 observables in $K_{\ell4}$, 18 in the scalar formfactors
and 11 in the vectorformfactors. This means a total of 76 quantities. We
 found a total of
35 relations between these. Most of these had been noticed earlier but
 we did find
several new ones. We have presented the relations in a form as simple as we
found but given the total number they can be rewritten in many equivalent forms.

These are relations that should allow independent determinations
of combinations of the NNLO LECs in ChPT. For the vector formfactors this was
already done in \cite{BT2,BT3} and partly for the $\pi\pi$, $\pi K$
 system \cite{BDT1,BDT2,KM}
and scalar formfactors \cite{BD}. Here we studied in detail
 the relations for the $\pi\pi$,
$\pi K$ scattering and $K_{\ell4}$ since for these cases enough experimental
 and/or dispersion
theory results exist. Fig.~\ref{figfullrelations} is a summary
of the numerical relations.

 The resulting picture is that ChPT at NNLO works in most
cases but there are some problems. The $\pi\pi$ system alone
 is working well,
the $\pi K$ system alone works satisfactorily but with some problems. The same
can be said for the combinations of both systems. A common part
in these two cases is the presence
of $a^-_3$. Comparing $\pi K$ scattering and $K_{\ell4}$ there
 is a clear contradiction.
In fact, both sides of the relation seem to be difficult to explain within ChPT.
It was already noticed in \cite{ABT3} that getting such a
 large negative curvature in $K_{\ell4}$ was difficult. 
It should be noted that none
of the quantities involved in the
tested relations was used as input for the fit of the NLO LECs in \cite{ABT4}.

\section*{Acknowledgements}

IJ gratefully acknowledges an Early Stage Researcher position supported by the
EU-RTN Programme, Contract No. MRTN--CT-2006-035482, (Flavianet)
This work is supported in part by the European Commission RTN network,
Contract MRTN-CT-2006-035482  (FLAVIAnet), 
European Community-Research Infrastructure
Integrating Activity
``Study of Strongly Interacting Matter'' (HadronPhysics2, Grant Agreement
n. 227431)
and the Swedish Research Council.

\appendix
\section{Relation between threshold and subtreshold parameters}

For completeness we quote here the relations between the threshold
and subthreshold parameters for the tree level part, i.e. that
analytic dependence on $s$, $t$ and $u$.
The $\pi\pi$ ones can also be found in \cite{BCEGS1} and \cite{ABT3}.
For the $\pi K$ case we have already used the relation (\ref{piKrel})
\ba
\pi   a^0_0 &=& 
          6 b_5 
       +   b_4 + (3/2) b_3 + (3/8) b_2 + (5/32) b_1
\,,\nonumber\\
\pi   b^0_0 &=&
         - 2 b_6 + 18 b_5 
       +   3 b_4 + 3 b_3 + (1/4) b_2 
\,,\nonumber\\
\pi   a^2_0 &=&
        b_4 + (1/16) b_1
\,,\nonumber\\
\pi   b^2_0 &=&
        - 2 b_6 
       +   3 b_4 - (1/8) b_2
\,,\nonumber\\
\pi   a^1_1 &=&
         ( 2/3) b_6 
       +  ( 1/3) b_4 + (1/24) b_2 
\,,\nonumber\\
\pi   b^1_1 &=&
         ( 4/3) b_6
       +  ( 1/2) b_4 - (1/6) b_3 
\,,\nonumber\\
\pi   a^0_2 &=&
         ( 16/15) b_6 
       +  ( 7/30) b_4 + (1/30) b_3 
\,,\nonumber\\
\pi   b^0_2 &=&
         ( 17/15) b_6 - (1/5) b_5 
\,,\nonumber\\
 \pi  a^2_2 &=&
         ( 4/15) b_6 
       +  ( 1/30) b_4 + (1/30) b_3 
\,,\nonumber\\
 \pi  b^2_2 &=&
         ( 1/3) b_6 -( 1/5 ) b_5
\,,\nonumber\\
 \pi  a^1_3 &=&
         ( 1/35) b_6 + (1/35) b_5 
\,.
\ea
\ba
\pi \left(\rho+1\right)  a^-_0 &=&
        24 \rho^3  c^-_{01}
       + ( 3/2) \rho   c^-_{00}
\,,\nonumber\\
 \pi\left(\rho+1\right)  b^-_0 &=&
       ( 36 \rho + 24 \rho^2 + 36 \rho^3 )  c^-_{01}
        - 3 \rho                         c^-_{10}
      + ( (3/4) \rho^{-1} + (3/4) \rho )         c^-_{00}
\,,\nonumber\\
\pi \left(\rho+1\right)  a^-_1 &=&
        12 \rho^2  c^-_{01}
       +  \rho      c^-_{10}
       + ( 1/4 )       c^-_{00}
\,,\nonumber\\
\pi \left(\rho+1\right)  b^-_1 &=&
         ( 12 - 6 \rho + 12 \rho^2 )          c^-_{01}
       +  (  - 1/2 + (1/2) \rho^{-1} + (1/2) \rho )c^-_{10}
       +    -( 1/8) \rho^{-1}                  c^-_{00}
\,,\nonumber\\
 \pi \left(\rho+1\right) a^-_2 &=&
         ( 24/5) \rho c^-_{01}
       +  ( 1/5 )      c^-_{10}
\,,\nonumber\\
 \pi\left(\rho+1\right)  b^-_2 &=&
         (  - (12/5) + (12/5) \rho^{-1} + (12/5) \rho )c^-_{01}
         -( 1/10) \rho^{-1}                    c^-_{10}
\,,\nonumber\\
 \pi\left(\rho+1\right)  a^-_3 &=&
        ( 24/35 ) c^-_{01}
\,,\nonumber\\
\pi\left(\rho+1\right)   a^+_0 &=&
        6 \rho^2  c^+_{01}
       + ( 3/8 )      c^+_{00}
\,,\nonumber\\
\pi\left(\rho+1\right)   b^+_0 &=&
         - 12 \rho^2          c^+_{11}
       +  ( 6 + 3 \rho + 6 \rho^2 )c^+_{01}
         - (3/4 )               c^+_{10}
         -( 3/16) \rho^{-1}   c^+_{00}\,.
\,,\nonumber\\
\pi\left(\rho+1\right)   a^+_1 &=&
        4 \rho^2  c^+_{11} 
       +2 \rho    c^+_{01} 
       +( 1/4 )      c^+_{10} 
\,,\nonumber\\
\pi\left(\rho+1\right)   b^+_1 &=&
        ( 4 - 2 \rho + 4 \rho^2 ) c^+_{11}
       + ( \rho^{-1} + \rho )      c^+_{01}
         -                   c^+_{20}
         - (1/8) \rho^{-1}       c^+_{10}
\,,\nonumber\\
\pi \left(\rho+1\right)  a^+_2 &=&
         ( 8/5) \rho  c^+_{11}
       +  ( 1/5 )     c^+_{01}
       +  ( 1/5 )     c^+_{20}
\,,\nonumber\\
 \pi\left(\rho+1\right)  b^+_2 &=&
        (  - (2/5) + (4/5) \rho^{-1} + (4/5) \rho ) c^+_{11}
         -( 1/10) \rho^{-1}                  c^+_{01}
         - (6/5 )                            c^+_{30}
         - (1/10) \rho^{-1}                  c^+_{20}
\,,\nonumber\\
 \pi \left(\rho+1\right) a^+_3 &=&
        ( 6/35 ) c^+_{11}
       + ( 6/35 ) c^+_{30}
\ea
It can be checked that these satisfy the relations given in Sects.~\ref{pipi},
\ref{piK} and \ref{pipipiK}.


\begin{thebibliography}{99}

\bibitem{Weinberg0}
S.~Weinberg,
Physica A {\bf 96}, 327 (1979).

\bibitem{GL0}
J.~Gasser and H.~Leutwyler,
Annals Phys.\  {\bf 158} (1984) 142.

\bibitem{GL1}
J.~Gasser and H.~Leutwyler,
Nucl.\ Phys.\  B {\bf 250} (1985) 465.

\bibitem{review}
V.~Bernard and U.~G.~Meissner,
Ann.\ Rev.\ Nucl.\ Part.\ Sci.\  {\bf 57} (2007) 33
[hep-ph/0611231].

\bibitem{reviewp6}
J.~Bijnens,
Prog.\ Part.\ Nucl.\ Phys.\  {\bf 58} (2007) 521
[hep-ph/0604043].

\bibitem{EFT09hans}
J.~Bijnens,
PoS {\bf EFT09} (2009) 022 [arXiv:0904.3713 [hep-ph]].

\bibitem{EFT09ilaria}
J.~Bijnens and I.~Jemos,
PoS {\bf EFT09} (2009) 032 [arXiv:0904.3705 [hep-ph]].

\bibitem{BCE1}
J.~Bijnens, G.~Colangelo and G.~Ecker,
{{\em J. High Energy Phys.}} {\bf 9902} (1999) 020
[hep-ph/9902437].

\bibitem{BCE2}
J.~Bijnens, G.~Colangelo and G.~Ecker,
{\em Annals Phys.}\  {\bf  280} (2000) 100
[hep-ph/9907333].


\bibitem{CGL}
G.~Colangelo, J.~Gasser and H.~Leutwyler,
{\em Nucl. Phys.} B {\bf  603} (2001) 125
[hep-ph/0103088].

\bibitem{ABT4}
G.~Amor\'os {\it et al.}, 
{\em Nucl. Phys.} B {\bf  602} (2001) 87
[hep-ph/0101127].

\bibitem{BDT1}
J.~Bijnens, P.~Dhonte and P.~Talavera,
{{\em J. High Energy Phys.}} {\bf  0401} (2004) 050
[hep-ph/0401039].

\bibitem{BCEGS1}
J.~Bijnens {\it et al.}, 
{\em Phys. Lett.}\ B {\bf  374} (1996) 210
[hep-ph/9511397];
%
{\em Nucl. Phys.} B {\bf  508} (1997) 263
[Erratum-ibid.\ B {\bf  517} (1998) 639]
[hep-ph/9707291].

\bibitem{BDM}
P.~Buettiker, S.~Descotes-Genon and B.~Moussallam,
{\em Eur.\ Phys.\ J.}\ C {\bf 33} (2004) 409
[hep-ph/0310283].

\bibitem{BDT2}
J.~Bijnens, P.~Dhonte and P.~Talavera,
{\em J. High Energy Phys.} {\bf 0405} (2004) 036
[hep-ph/0404150].

\bibitem{KM}
K.~Kampf and B.~Moussallam,
Eur.\ Phys.\ J.\  C {\bf 47} (2006) 723
[hep-ph/0604125].

\bibitem{Pislak2}
S.~Pislak {\it et al.}  [BNL-E865], 
{\em Phys. Rev.}\ D {\bf  67} (2003) 072004
[hep-ex/0301040].

\bibitem{NA48kl4}
J.~R.~Batley {\it et al.}  [NA48/2 Collaboration],
Eur.\ Phys.\ J.\  C {\bf 54} (2008) 411.

\bibitem{Bijnens:1994me}
  J.~Bijnens, G.~Colangelo, G.~Ecker and J.~Gasser,
  hep-ph/9411311.

\bibitem{Amoros:1999mg}
  G.~Amor\'os and J.~Bijnens,
  J.\ Phys.\ G {\bf 25} (1999) 1607
  [hep-ph/9902463].

\bibitem{ABT3}
G.~Amor\'os {\it et al.}, 
{\em Phys. Lett.}\ B {\bf 480} (2000) 71
[hep-ph/9912398];
{\em Nucl. Phys.} B {\bf  585} (2000) 293
[Erratum-ibid.\ B {\bf  598} (2001) 665]
[hep-ph/0003258].

\bibitem{BCG}
  J.~Bijnens, G.~Colangelo and J.~Gasser,
  Nucl.\ Phys.\  B {\bf 427} (1994) 427
  [arXiv:hep-ph/9403390].

\bibitem{BG2}
J.~Bijnens and K.~Ghorbani,
{\em J. High Energy Phys.} {\bf 0711} (2007) 030
[arXiv:0709.0230 [hep-ph]].

\bibitem{BD}
J.~Bijnens and P.~Dhonte,
{{\em J. High Energy Phys.}} {\bf  0310} (2003) 061
[hep-ph/0307044].

\bibitem{ABT1}
G.~Amor\'os {\it et al.}, 
{\em Nucl. Phys.} B {\bf  568} (2000) 319
[hep-ph/9907264].

\bibitem{GK}
E.~Golowich and J.~Kambor,
{\em Phys. Rev.}\ D {\bf 58} (1998) 036004
[hep-ph/9710214].

\bibitem{GL2}
  J.~Gasser and H.~Leutwyler,
  Nucl.\ Phys.\  B {\bf 250} (1985) 517.

\bibitem{BT3}
J.~Bijnens and P.~Talavera,
{\em Nucl. Phys.} B {\bf  669} (2003) 341
[hep-ph/0303103].

\bibitem{BT2}
J.~Bijnens and P.~Talavera,
{{\em J. High Energy Phys.}} {\bf  0203} (2002) 046
[hep-ph/0203049].

\bibitem{PS1}
P.~Post and K.~Schilcher,
%
{\em Phys. Rev. Lett.}\  {\bf 79} (1997) 4088
[hep-ph/9701422];
{\em Nucl. Phys.} B {\bf  599} (2001) 30
[hep-ph/0007095].


\end{thebibliography}
\end{document}